\documentclass{article}

\usepackage{arxiv}

\usepackage[utf8]{inputenc} % allow utf-8 input
\usepackage[T1]{fontenc}    % use 8-bit T1 fonts
\usepackage{hyperref}       % hyperlinks
\usepackage{url}            % simple URL typesetting
\usepackage{booktabs}       % professional-quality tables
\usepackage{amsfonts}       % blackboard math symbols
\usepackage{nicefrac}       % compact symbols for 1/2, etc.
\usepackage{microtype}      % microtypography
\usepackage{lipsum}

\usepackage{graphicx}
\usepackage{changepage}
\usepackage{longtable}
\usepackage{amssymb}
\usepackage{amsmath}

\title{Correlation Patterns in Foreign Exchange Markets}% Force line breaks with \\
\author{
  Lasko Basnarkov \\
  Faculty of Computer Science and Engineering\\
 SS. Cyril and Methodius University\\
  P.O. Box 393, 1000 Skopje, Macedonia \\
  \texttt{lasko.basnarkov@finki.ukim.mk} \\
   \And
 Viktor Stojkoski \\
  Center for Computer Science and Engineering\\
Macedonian Academy of Sciences and Arts\\
  P.O. Box 428, 1000 Skopje, Macedonia \\
  \texttt{vstojkoski@manu.edu.mk} \\
     \And
 Zoran Utkovski \\
  Fraunhofer Heinrich Hertz Institute\\
  Einsteinufer 37, 10587, Berlin, Germany \\
  \texttt{zoran.utkovski@hhi.fraunhofer.de} \\
       \And
 Ljupco Kocarev \\
  Center for Computer Science and Engineering\\
Macedonian Academy of Sciences and Arts\\
  P.O. Box 428, 1000 Skopje, Macedonia \\
  \texttt{lkocarev@manu.edu.mk} \\ \\
}

\begin{document}
\maketitle

\begin{abstract}
The value of an asset in a financial market is given in terms of another asset known as numeraire. The dynamics of the value is non-stationary and hence, to quantify the relationships between different assets, one requires convenient measures such as the means and covariances of the respective log returns. Here, we develop transformation equations for these means and covariances when one changes the numeraire. The results are verified by a thorough empirical analysis capturing the dynamics of numerous assets in a foreign exchange market. We show that the partial correlations between pairs of assets are invariant under the change of the numeraire. This observable quantifies the relationship between two assets, while the influence of the rest is removed. As such the partial correlations uncover intriguing observations which may not be easily noticed in the ordinary correlation analysis.
\end{abstract}

% keywords can be removed

\section{Introduction}
Financial markets` data has traditionally challenged economists, mathematicians, and statisticians. However, in recent years, physicists have joined in providing answers on different features of these data which resulted in the emergent field of econophysics~\cite{mantegna1999introduction}. 

A particularly interesting feature of financial markets data is the, not yet fully explored, complex interdependent dynamics of the prices between various assets, which is most often modeled through the empirical correlations. This issue has been of specific value to the econophysics community and its members have provided numerous contributions addressing the relationships between stocks, market indexes and currencies~\cite{mantegna1999hierarchical, laloux1999noise, plerou1999universal, mizuno2006correlation, naylor2007topology, drozdz2007world, keskin2011topology, vodenska2016community, mai2018currency}. 

In a foreign exchange market, the pairwise correlations between assets constitute a matrix. To understand the significant relationships in this matrix, one usually constructs a network by linking the strongly correlated pairs. The pairs are usually chosen via standard algorithms, such as Minimum Spanning Tree (MST)~\cite{mantegna1999hierarchical} or Planar Maximally Filtered Graph (PMFG)~\cite{tumminello2005tool, tumminello2007correlation}. It has been acknowledged that the correlation matrix and hence the networks generated by MST and PMFG, in general depend on the choice of the base asset, formally known as \textit{numeraire}, used to measure the value of all other assets. Moreover, changing the numeraire may also distort the plausibility of several economic theories. For instance, the key theory of exchange rates, the Purchasing Power Parity, relies on numeraire invariance, i.e., different numeraires should yield same conclusions. However, a growing body of literature negates the empirical validity of this property in foreign exchange markets~\cite{papell2001choice,hovanov2004computing,kremer2018economic}. In spite of these observations, the exact implications created by substituting the numeraire for another asset are yet to be determined.

Relating variables measured in different reference frames has a long history in physics. For example, the famous Galilean~\cite{galilei2011discorsi} and Lorentz~\cite{lorentz1937electromagnetic} transformations are used for establishing correspondences between kinematic variables in classical and relativistic scenarios, respectively. Evidently, the role of reference frame in finance is played by the numeraire. In stock markets the natural choice of numeraire is the domestic currency, whereas at a foreign exchange different assets may fit for this purpose. In studies examining the properties of the latter markets, usually the most traded asset is taken as numeraire, while another less liquid asset is used to compare the results~\cite{naylor2007topology, keskin2011topology}. 

Motivated by these observations, here we derive the transformation equations for the key statistical quantities (means and covariances) when the numeraire is changed. As an application of these transformations, we discuss the choice of numeraire used for valuing the performance of a portfolio of assets, and provide arguments that the most appropriate numeraire is the local currency of the investor. More importantly, we give a thorough empirical analysis which not only confirms our findings, but also provides a detailed overview on the influence of the numeraire choice on the resulting correlations. In particular, we show that there is a typical pattern where certain correlations which are significant in one numeraire become insignificant in another.  

Finally, we also comment on a numeraire invariant measure for capturing the correlation between assets. In fact, we find that one such measure is the partial correlation -- the quantification of the linear relationship between two assets while controlling for the potential effect of all other assets in the market. As such, a network constructed via the partial correlations may provide a more adequate illustration on the pairwise relationships in a foreign exchange market.

The rest of the paper is organized as follows. In Section~\ref{sec:Theory} we begin by providing the theoretical background and analysis of the problem of numeraire change. Section~\ref{sec:Data} describes the data used for empirical verification of the developed theory. In Section~\ref{sec:Results} we present the empirical results and discuss the implications created by substituting the numeraire. Finally, Section~\ref{sec:Conclusions} summarizes our findings.

\section{Theoretical analysis}
\label{sec:Theory}

The dynamics of exchange rates is in general non-stationary, and hence non-ergodic. This implies that one cannot exploit the raw values to examine the relationships between different assets. A convenient way to circumvent this problem is to instead focus on the logarithmic returns of exchange rates, which are known for their stationary dynamics~\cite{peters2016evaluating}.

We assume that the distribution of these log returns is multivariate Gaussian, which is a plausible assumption under the efficient market hypothesis~\cite{roll1984simple}, even though there are observations suggesting otherwise~\cite{mandelbrot1967variation, drozdz2010foreign, corlu2015modelling}. It is widely known that the Gaussian distribution is fully defined by its mean and covariance structure, which in turn is dependent on the chosen numeraire. This is of particular relevance in a foreign exchange market, as in it the value of any asset is given as a ratio with respect to another asset, i.e. the exchange rate. Thus, there is no single value of an asset, but many relative prices expressed in terms of the other assets. Once one selects a numeraire, even a random one, and hence determines the value of some asset in the market, in absence of arbitrage the values of the remaining ones are also determined. This is analogous to the assignment of a voltage to some node, which uniquely determines the voltages of all other nodes in an electrical circuit. In the following we analytically evaluate the consequences of changing numeraire on the values of the means, covariances and correlations in an arbitrary foreign exchange market.

%Moreover, in under the efficient market hypothesis, consecutive returns of a single time series are nearly uncorrelated \cite{roll1984simple} and can be conveniently modeled by a Gaussian, i.e. normal distribution. Although the histograms of the log returns show significant departure from the normal distribution~\cite{mandelbrot1967variation, corlu2015modelling}, especially at the tails, this distribution has been dominant in modeling the statistics of returns due to its mathematical tractability. Among other conveniences, even when dealing with multiple random variables, the respective multivariate Gaussian distribution is completely defined by the marginal means, variances and covariances between each pair~\cite{fama1976foundations}. This feature of the normal distribution appears also very attractive in construction of optimal portfolios~\cite{markowitz1952portfolio}.

\subsection{Mean and covariance transformations}

We begin by assuming that the observations arrive discretely at moments $1, 2, \dots$, and randomly assign values to one of the currencies $U_1, U_2, \dots$. The values of the remaining currencies are then determined from the existing exchange rates. This indicates that the log return $x^u$ of any currency $X$ given in the numeraire $U$ at a particular moment $n$ would be
\begin{equation}
    x^u = \log\left(\frac{X_{n}}{U_{n}}\right) - \log\left(\frac{X_{n-1}}{U_{n-1}}\right).
\label{eq:log_return_def}
\end{equation}
The return $x^w$ of $X$ in another numeraire $W$ is simple linear combination of the returns
\begin{equation}
    x^w = \log\left( \frac{X_{n}}{U_{n}}\frac{U_{n}}{W_{n}}\right) - \log\left(\frac{X_{n-1}}{U_{n-1}}\frac{U_{n-1}}{W_{n-1}}\right) =  x^u - w^u.
    \label{eq:log_ret_transform}
\end{equation}
Since we are working with ergodic observables, we can apply the averaging rules, and conclude that the change of numeraire results in linear transformation of the mean of log returns as well,
\begin{equation}
\langle x^w \rangle = \langle x^u \rangle - \langle w^u \rangle,
\label{eq:expectation_transform}
\end{equation}
where the brackets denote averages. The covariance between the log returns of currencies $X$ and $Y$ in the first numeraire $U$ is defined as
\begin{equation}
C_{x,y}^u = \langle (x^u - \langle x^u \rangle )( y^u - \langle y^u\rangle) \rangle.    
\label{eq:cov_definition}
\end{equation} 
By using the return transformations~(\ref{eq:log_ret_transform}) in the definition of the covariance~(\ref{eq:cov_definition}), after simple algebra one can easily show that the covariance in the second base is
\begin{equation}
C_{x,y}^w = C_{x,y}^u - C_{x,w}^u - C_{y,w}^u + C_{w,w}^u.
\label{eq:Cov_transform}
\end{equation}
The last expression is transformation of the covariance from one numeraire to another, and resembles those used in kinematics. Furthermore, from~(\ref{eq:Cov_transform}) we can directly extract the Pearson correlation coefficient, $r^w_{x,y} = C^w_{x,y}/\sqrt{C^w_{x,x}C^w_{y,y}}$, as
\begin{equation}
r_{x,y}^w = \frac{C_{x,y}^u - C_{x,w}^u - C_{y,w}^u + C_{w,w}^u}{\sqrt{\left(C_{x,x}^u - 2C_{x,w}^u + C_{w,w}^u\right)\left(C_{y,y}^u - 2C_{y,w}^u + C_{w,w}^u\right)}}.
\label{eq:Pearson_coeff_transf}
\end{equation} 

We note that a special case of the relationship~(\ref{eq:Cov_transform}) is the variance transform $C_{x,x}^w = C_{x,x}^u - 2C_{x,w}^u + C_{w,w}^u$, which has relevance in calculation of covariances when the three necessary variances are known~\cite{campa1998forecasting, lopez2000implied}. This relationship is widely used in determining the covariance between two assets from the corresponding values of their implied variances, which in turn are estimated from the respective option prices~\cite{buss2012measuring, symitsi2018covariance}.

\subsection{Variance of portfolio}

Modern portfolio theory states that the optimal portfolio is achieved by diversifying investments in a combination which leads to the lowest variance for given mean. It is evident that, in this aspect, changing the numeraire also modifies the parameter values of the portfolio -- its expected return and variance. To find the respective transformation that is needed when one changes the numeraire, we consider a portfolio of assets, $X_1, X_2, , \ldots, X_N$, which under base $U$ has value $X^u = \sum_{i=1}^N \alpha_i (X_i/U)$, where $\alpha_i$ is the share of investments in $X_i$. Consequently, its return in units of $U$ is
\begin{equation}
    x^u = \sum_{i=1}^N \alpha_i x_i^u.
    \label{eq:Portfolio_return}
\end{equation}
The portfolio return changes in the same manner as in equation~(\ref{eq:log_return_def}), and its value in the numeraire $W$ is
\begin{equation}
x^w = \sum_{i=1}^N \alpha_i (x_i^u - w^u) = x^u - w^u.
\label{eq:basket_transform}
\end{equation}
%We remind that the last expression is due to the normalization $\sum \alpha_i = 1$. Clearly, such simple expression was obtained due to the linear transformation of the log returns. 
Going further one can obtain the variance of the portfolio in numeraire $U$ as
\begin{equation}
C_{x,x}^u= \sum_{i=1}^N \alpha_i^2 C_{x_i,x_i}^u + \sum_{i\neq j} \alpha_i \alpha_j C_{x_i,x_j}^u.
\label{eq:portfolio_var_first}
\end{equation}
After expanding the terms appearing in the transformation of the portfolio return~(\ref{eq:basket_transform}) into the definition of the covariance~(\ref{eq:cov_definition}) and some rearrangements we find that
\begin{equation}
C_{x,x}^w = \sum_{i=1}^N \alpha_i^2 C_{x_i,x_i}^u + \sum_{\substack{i,j = 1\\i\neq j}}^N \alpha_i \alpha_j C_{x_i,x_j}^u - 2\sum_{i=1}^N \alpha_i C_{x_i,w}^u + C_{w,w}^u.
\end{equation}
This is compactly written as 
\begin{equation}
C_{x,x}^w = C_{x,x}^u - C_{x,w}^u + C_{w,w}^u,
\end{equation}
by defining $C_{x,w}^u = 2\sum_{i=1}^N \alpha_i C_{x_i,w}^u$ as the covariance between the portfolio and the second numeraire $W$ observed from the point of view of the first one $U$. %It appears that even when the currencies of the portfolio are uncorrelated with the second numeraire the variance in the two bases differ by the amount of the variance of the second numeraire as observed from the first one.

Similarly to the twin paradox in special relativity~\cite{einstein1905electrodynamics}, the dependence of the variance on the base currency poses a dilemma about which numeraire should be used as reference frame. To resolve this issue, let us assume that a particular investor is in a country with currency $U$. In order to finance a portfolio with starting value $X(s)$ the agent should borrow the same amount in currency $U$ from a local bank, and return it later at the final moment $f$, again in the same currency. Furthermore, suppose that the bank interest rate is zero and that the portfolio is valued at one currency unit $U$. Then, the initial and final wealth of the investor are correspondingly $0$ and $X^u(f) - U$. The variance of the portfolio is given by equation~(\ref{eq:portfolio_var_first}). 

The same investor should observe the variance expressed in the numeraire $W$ only if it is initially financed by a credit denominated in that asset. This follows by observing the initial credit $U/W(s)$, as valued in the second asset, and then purchasing the portfolio. In particular, by assuming negligible transaction costs (for simplicity), the value of the portfolio at purchase is
\begin{equation}
X^w(s) = \sum_{i=1}^n \alpha_i \frac{X_i(s)}{W(s)} = \sum_{i=1}^n \alpha_i \frac{X_i(s)}{U} \frac{U}{W(s)},
\end{equation}
in $W$ currency units, with the same expression holding at the closing time $f$. If one substitutes the final and starting moments $f$ and $s$, with $n$ and $n-1$ respectively, the return on the portfolio is given as in~(\ref{eq:basket_transform}), i.e. $x^w = x^u - u^w = x^u + w^u$. This means that the log return of the investments is in one part due to the portfolio return $x^u$ and in another one due to the exchange rate return $w^u$ between $W$ and $U$. On the other hand, it is obvious that the return of the borrowing cancels the exchange rate contribution of the return in investments since its value is $-w^u$. Consequently, the return on the portfolio and respectively its variance is identical in both cases, which means that it might be more convenient for the investor to use its domestic currency as numeraire.

\subsection{Partial correlation}

%The correlation between two random variables can be either a result of mutual interactions or due to common external factors.
To account for the possibility that other variables drive the relationship between the two variables under study, one calculates the respective partial correlation~\cite{mai2018currency, kenett2010dominating, kenett2015partial}. For the case of foreign exchange markets, this translates to the correlation between the log return of two assets controlled for the return of all other assets exchanged in the market. To examine the numeraire dependence of this quantity, let us first define the set $\mathbb{R}$ containing all possible returns of all assets in the foreign exchange as valued in \textit{every} possible numeraire. Then, the partial correlation between the returns of assets $X$ and $Y$ valued in numeraire $U$ reads
\begin{equation}
\rho^u_{x,y}= \langle \left(x^u - \langle x^u \rangle \right) \left(y^u - \langle y^u \rangle \right) |\mathbb{R}\backslash\{x^u, y^u\}\rangle.
\end{equation}
The partial correlation with numeraire $W$ is defined in the same way, only conditioned on $\mathbb{R}\backslash\{x^w, y^w\}$.
Notice that the linear transformation between the returns of $x$ in $u$ and $w$, $x^w=x^u - w^u$, contains the constant term $w^u$ which is the sets of known variables in consideration of the two partial correlations $\rho^u_{x,y}$ and $\rho^w_{x,y}$. Combining this fact with the property that adding constants in two variables does not change their correlation, implies that we can write the latter partial correlation as
\begin{eqnarray}
\rho^w_{x,y} &=& \langle \left(x^w - \langle x^w \rangle \right) \left(y^w - \langle y^w \rangle \right)  |\mathbb{R}\backslash\{x^w, y^w\} \rangle = \nonumber \\
&=& \langle \left(x^u - w^u - \langle x^u - w^u \rangle \right) \left(y^u - w^u - \langle y^u - w^u \rangle \right) |\mathbb{R}\backslash\{x^u, y^u\} \rangle = \nonumber\\
    &=& \langle \left(x^u - \langle x^u \rangle \right) \left(y^u - \langle y^u \rangle \right)  |\mathbb{R}\backslash\{x^u, y^u\} \rangle.
\end{eqnarray}
%Thus, we have proved something that seems intuitively trivial: 
Hence, the partial correlation between the return of two assets controlled for the returns of all other assets traded in the market, is invariant on the numeraire. As a side note, we point out that such partial correlations are easily calculated through the precision matrix $\mathbf{P}$ which is the inverse of the correlation matrix $\mathbf{C}$, and by utilizing the following normalization
\begin{equation}
    \rho_{x,y} = -\frac{P_{x,y}}{\sqrt{P_{x,x} P_{y,y}}},
    \label{eq:Partial_corr}
\end{equation}
for each pair of assets $X$ and $Y$ \cite{lauritzen1996graphical}.

\section{\label{sec:Data}Data}

To verify the presented findings, we study market data provided by the Pacific Exchange Rate Service  Dataset\footnote{\url{http://fx.sauder.ubc.ca/data.html}} (PERS). PERS is a freely available dataset which provides daily values of 92 currencies and commodities priced in various numeraires. In the analysis, we opted to include commodities because combining them with currencies allows for capturing more general patterns in financial markets. In fact, as we will see later on, there appear direct relations between the currencies and commodities.

As a time frame we utilized the four year period between 2014 and 2017. Since calculation of the log returns requires data for two consecutive days, we excluded from the analysis all data points for which there was no consecutive data\footnote{For calculating Monday returns we used the respective asset price in Friday, whereas for days when there were holidays we took the value on the day before the holiday.}. Finally, from the analysis we excluded all currencies for which there were at least ten missing values and those assets that have constant exchange rates with the USD for at least five working days. We note that the considered period, the precious metals present in the dataset have around 30 missing values. Nevertheless, we kept them in the analysis as they may represent a significant driver in the returns of several currencies. Altogether, we ended up with $897$ log returns data for $48$ currencies and $5$ commodities. The names of the assets together with their abbreviations are listed in Table~\ref{tab:Abbreviations}.

\begin{table}[h!]
\caption{List of abbreviations of the studied assets}
  \begin{center}
    
   \label{tab:Abbreviations}
    \begin{tabular}{|l|l||l|l|} % <-- Alignments: 1st column left, 2nd middle and 3rd right, with vertical lines in between
       \hline 
        \multicolumn{4}{|c|}{\textbf{Currencies}} \\
      \hline 
     AED & U.A. Emirates dirham & KRW & Korean won\\ \hline
     ARS & Argentine peso & KWD & Kuwaiti dinar\\ \hline
     AUD & Australian dollar & LKR & Sri Lankan rupee\\ \hline
     BHD & Bahraini dinar & MAD & Moroccan dirham\\ \hline
     BRL & Brazilian real & MXN & Mexican peso\\ \hline
     CAD & Canadian dollar & MYR & Malaysian ringgit\\ \hline
     CHF & Swiss franc & NOK & Norwegian krone\\ \hline
     CLP & Chilean peso & NZD & New Zealand dollar\\ \hline
     CNY & Chinese renminbi & PEN & Peruvian nuevo sol\ \\ \hline
     COP & Colombian peso & PHP & Philippines peso\\ \hline
     CZK & Czech koruna & PLN & Polish zloty\\ \hline
     DKK & Danish krone & RUB & Russian ruble\\ \hline
     EUR & EURO & SAR & Saudi Arabian riyal\\ \hline
     GBP & British pound & SEK & Swedish krona\\ \hline
     HKD & Hong Kong dollar & SGD & Singapore dollar\\ \hline
     HNL & Honduran lempira & THB & Thailand baht\\ \hline
     HRK & Croatian kuna & TND & Tunisian dinar\\ \hline
     HUF & Hungarian forint & TRY & Turkish lira\\ \hline
     IDR & Indonesian rupiah & TWD & Taiwanese dollar\\ \hline
     ILS & Israeli new sheqel & USD & U.S. dollar\\ \hline
     INR & Indian rupee & UYU & Uruguayan peso \\ \hline
     ISK & Icelandic krona & VND & Vietnamese dong\\ \hline
     JMD & Jamaican dollar & XCD & East Caribbean dollar \\ \hline
     JPY & Japanese yen & ZAR & South African rand\\ \hline
     \hline      
     \multicolumn{4}{|c|}{\textbf{Commodities}} \\
     \hline      
	 XAG & Silver ounce & XCT & West Texas barrel\\ \hline
     XAU & Gold ounce & XPT & Platinum ounce\\ \hline
     XCB & Brent Crude barrel & \multicolumn{2}{|c|}{} \\
     \hline
    \end{tabular}
  \end{center}
\end{table}

We point out that the initial data was gathered with US dollar (USD) as numeraire. When calculating the asset values in other numeraires we utilized the triangular equivalence, i.e., a rate $X/Y$ was estimated from the known $X/Z$ and $Y/Z$ as $X/Y = (X/Z) / (Y/Z)$. This is a common approach~\cite{fenn2012dynamical}, even though small deviations from the real values known as triangular arbitrage, were found on shorter time scales. On longer scales, such as the daily ones, they are negligible~\cite{drozdz2010foreign}. Once all exchange rates were determined, by taking each of the 53 assets as numeraire, we calculated the daily log returns of the remaining 52 assets, and estimated the corresponding covariance and correlation matrices.
%In order to distinguish between correlation which is due to some mutual influence of the assets, from that obtained by chance, one usually applies statistical significance test. If the probability of random appearance of correlation larger than the one under test, is lower than some threshold, the result is considered as statistically significant. In significance tests for single correlation coefficient the threshold is usually taken to be 0.05, or 0.01. 
To test the significance of all terms in the correlation (covariance) matrices we implemented the Bonferroni correction~\cite{tumminello2011statistically, curme2015emergence}. Under this test, the level of significance is given by dividing the level of significance under single test with the number of elements in the correlation matrix. Accordingly, we took the level of significance to be $0.05 / (53 \cdot 53)$.

%When one applies it, the value of the corresponding threshold is obtained by dividing that for single test by the number of correlation matrix elements. Respectively, we have used as threshold for considering the correlation coefficient as statistically significant if the probability of its occurrence from two independent time series is smaller than $0.05 / (53 \cdot 53)$. For independent Gaussian time series, by using Fisher transform, one can obtain that correlations that pass such statistical significance test should exceed 0.1374 by absolute value.}

\section{\label{sec:Results}Results}

\subsection{Empirical validation of the transformations}

\begin{figure*}[t!]
\begin{adjustwidth}{-0.2in}{0in}
\includegraphics[width=15cm]{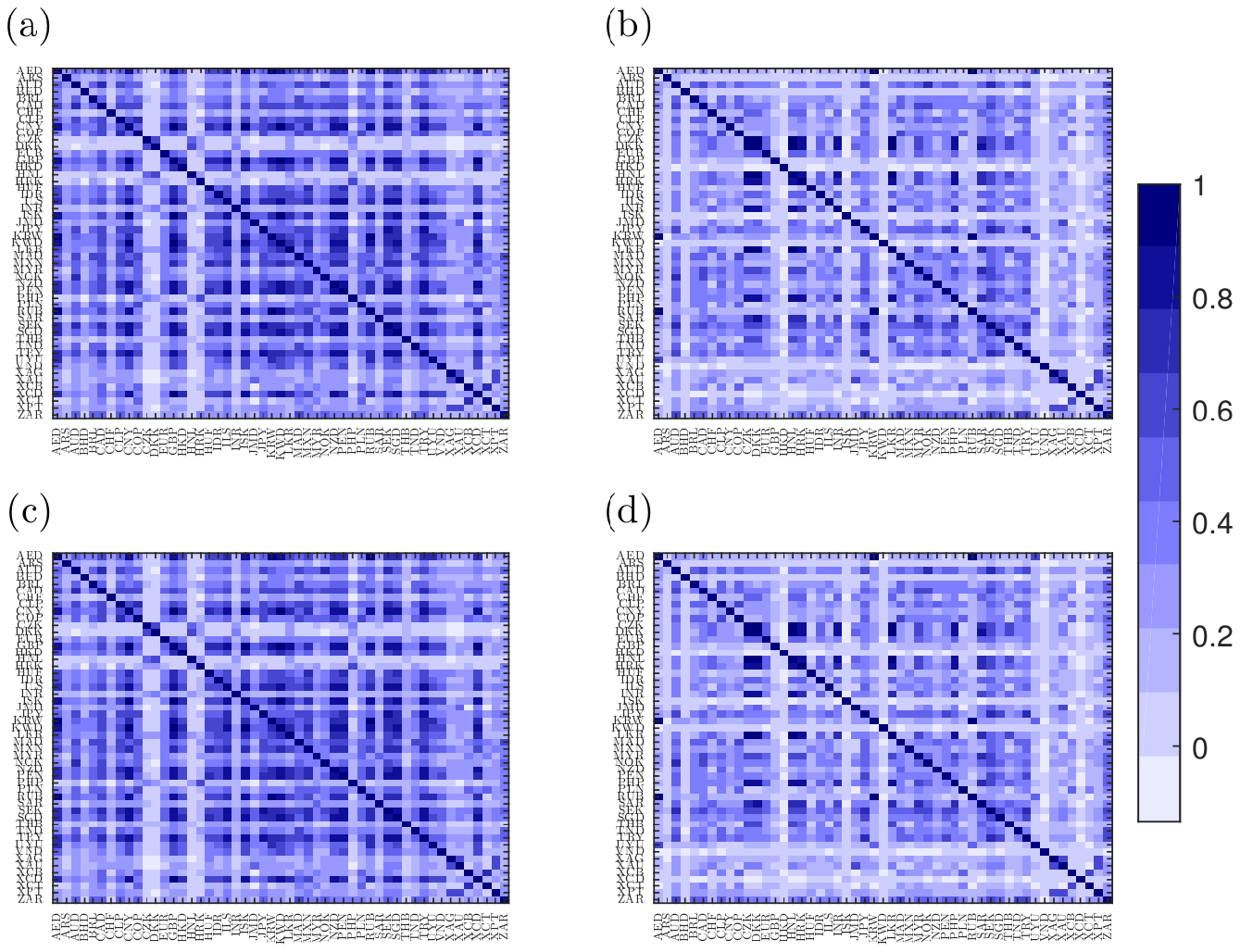}
\caption{\textbf{Correlation heat maps.} \textbf{(a)} Correlation estimated from data where the numeraire is EUR. \textbf{(b)} Same as \textbf{(a)} only the numeraire is USD \textbf{(c)} Correlations in EUR estimated from transforming the correlation matrix in USD as base currency \textbf{(d)} Same as \textbf{(c)} only the correlations are estimated by transforming the EUR matrix to USD.
\label{fig:correlation-matrix}}
\end{adjustwidth}
\end{figure*}

First, we verify the validity of the transformations developed in the previous section. For this purpose, in Fig.~\ref{fig:correlation-matrix} we plot heat maps for the empirical correlations between the studied assets as valued in EUR (Fig.~\ref{fig:correlation-matrix}a) and in USD (Fig.~\ref{fig:correlation-matrix}b), and their corresponding transformed correlations, from USD to EUR in Fig.~\ref{fig:correlation-matrix}c and from EUR to USD in Fig.~\ref{fig:correlation-matrix}d. One can easily notice the striking resemblance between these two types of correlations. In fact, the coefficient of determination, which quantifies the percentage of the variance in the empirical correlations explained with the transformed correlations, states that almost 100 percent of the variance in both the EUR matrix and USD matrix is explained with the transformations. The same conclusion holds for all other empirical correlations and transformation pairs.

\subsection{Implications of changing the numeraire}

Fig.~\ref{fig:correlation-matrix} also reveals the known fact that the numeraire ultimately determines the relationships in the correlation matrix~\cite{naylor2007topology,drozdz2007world,keskin2011topology}. While one typically uses the correlation matrix to build networks via MST or PMFG, and thus study the leading currencies and the clusters of mutually related assets, here instead we focus on the properties of the correlation matrix. Even though, this is not necessarily the same, we point out that the properties of the correlation matrix are those which uniquely determine the structure of the network.

For this purpose we turn our attention to the correlation matrix as estimated with USD as base currency. In the Appendix in Table~\ref{tab:Corr_USD} are shown the correlation coefficients obtained with this currency as numeraire, which are larger than 0.6. In order to extract the more significant ones, we apply filtering by considering only ones that exceed certain threshold.

First, when the threshold is 0.9, one can notice  that for this numeraire the strongest positive correlations appear between the currencies of oil exporting Middle Eastern countries, i.e. between AED, KWD and SAR. They are accompanied by a clique of European currencies, CZK, DKK, EUR, and HRK together with the North African MAD. By lowering the threshold to 0.8, HUF and the PLN join the European cluster, while taking an even lower threshold of 0.6, further adds NOK, SEK and TND to the group. At the same time, this results in emergence of an East Asian clique containing the KRW, SGD and TWD, and significance in the correlation between the Oceania pair of AUD and NZD. One can easily hypothesize that the existence of these relations has roots in the geographical proximity of the countries, and hence increased economic interdependence. Without going into further analysis of the other, weaker correlation patterns, we just note that the three major currencies CHF, GBP and JPY in these USD-based calculations are close to the European cluster. The last currency also has very strong correlation with the gold.

If one uses another currency as numeraire, for instance EUR, remarkable differences in the correlation coefficients from those observed in the USD-based calculations are identified. By considering the correlations larger than 0.9, given in the Appendix in Table \ref{tab:Corr_EUR}, one can notice that the Middle East currency trio AED, KWD and SAR is present again. Also a new clique of correlations exceeding 0.9 arises and which consists of CNY, JMD, HKD, LKR and USD. In addition, the correlation between XCD and HKD is also above 0.9. By decreasing the threshold, the later two cliques coalesce, while simultaneously attracting new Asian currencies, in particular, INR, PHP, THB and TWD together with the Latin American PEN, to the cluster. Further lowering the threshold to 0.6 results in inclusion of new members to the big cluster. The new members are the Asian IDR, ILS, KRW, MYR, SGD, and VND, the Latin American currencies CLP and HNL and the CAD. Also, under a lowered threshold one notices the correlation between the Oceania couple AUD and NZD and the joining of MAD to the USD-based mega cluster. It is worth to point out that the MAD was discovered to be rather correlated to the European currencies when observed from the USD as numeraire. 

From this correlation analysis one notices that the European cluster appears within USD-based calculations, while the currencies that closely track the dynamics of the US dollar can be detected when EUR is the numeraire. From this observation one could further conjecture a test for determining the strong correlations between two assets $X$ and $Y$ based on two conditions. First, they should be highly correlated in various numeraires, and second, when one of them is used as numeraire (for example $X$), the correlations of the other, $Y$, with the remaining assets should become smaller. Loosely speaking, in base currency $X$ the correlations of $Y$ with the others are masked. If any other currency $Z$ is correlated to both $X$ and $Y$, then the correlation between $X$ and $Y$ in numeraire $Z$, with the other currencies $U$, $r_{y,u}^z$ and $r_{x,u}^z$, will be small as well. Thus, using the CZK, DKK, or HRK as numeraire produces similar results as EUR -- weak correlations of the other currencies from this European group with the rest, while keeping the Middle East and USD clusters of highly correlated currencies. We also found that the masking phenomenon has an effect even on the currencies which could not be easily associated to some cluster. For example, when considered EUR as base currency, the CHF, GBP, and JPY seem to be more correlated to the currencies of the USD cluster, while as seen from the USD perspective they appear closer to the European group.

To confirm the masking phenomenon we also considered using SAR as numeraire, which is the currency of the largest country of the strongly correlated Middle East trio. The respective Pearson correlation coefficients exceeding 0.6 are provided in the Appendix in Table \ref{tab:Corr_SAR}. When filtering the correlations larger than 0.9 one discovers only the European clique, consisting of the CZK, DKK, EUR, HRK and MAD, while a USD-based cluster containing CNY, HKD, JMD, and LKR, appears when the threshold is lowered to 0.8. Again, by considering even weaker correlations, other currencies join to the respective groups, until they merge into one big cluster. For this numeraire we found an interesting exception of the masking phenomenon, i.e., we discovered a high correlation of 0.82 between AED and BHD which was previously missing. This suggests that BHD should belong to a cluster where AED is. As we pointed out previously, such strong correlation should also be observed from other numeraires like the EUR or USD. In fact, the masking should result in weak correlation between AED and BHD when the numeraire is SAR, since,  as discovered from the view of the two previous numeraires, SAR is rather closely related to AED. Nevertheless, the results show the opposite. The interpretation behind this finding may be due to the fact both AED and BHD are currencies of oil exporting countries, or due to the economic relations with other countries.  In what follows we will offer further indications for the plausibility of the second interpretation, since a very strong but \textit{negative} partial correlation between SAR and BHD appears. Lastly, we note that using SAR as numeraire pushes the CHF, GBP, and JPY closer to EUR, whereas the previously observed strong correlation of JPY with the gold persists since JPY does not belong to either of the USD and EUR groups.

\subsection{Robustness of correlated pairs}

As a means to understand the influence of the choice of numeraire on the correlations between other currencies that probably have more independent dynamics, for each asset $X$ we determined the maximal correlation under numeraire $U$, i.e.,
\begin{equation}
    r_{x,\max}^u = \max_z \left\{r_{x,z}^u \right\},
\end{equation}
by searching for it over all other assets $Z$. The asset $Y$ for which the strongest correlation $r_{x,\max}^u$ is obtained can be considered as the most similar to $X$ when the numeraire is $U$. Obviously, this similarity relationship is asymmetrical, and by substituting the numeraire, the most similar asset to $X$ could change as well. In this regard, by taking all numeraires under consideration, for each asset we estimate the number of occurrences of most similar assets. This allows us to effectively reduce the impact of numeraire in determining the similarity and thus develop a robust measure for the significance of the similarity. Clearly,  higher number of such occurrences does not imply more pronounced similarity between two assets, as it does not quantify the magnitude of it. It only offers a robust way to identify the asset which most often appears to be ``closest'' in terms of correlation. Even though, this approach may appear more computationally demanding than the standardly used MST and PMFG, it offers a more comprehensive overview of quantifying the most similar asset as it considers all possible numeraires.
%We remark that this analysis is not very rigorous because we had not considered the strengths of the correlation coefficients in the selection of the closest asset. Also, because EUR and USD have many closely related currencies with them, their perspective has more weight because their satellite currencies show similar results. This approach seems to complement the MST and PMFG, although being more computationally demanding

The results can be seen in Table~\ref{tab:Close_assets} where we report each asset $X$ alongside its most similar asset $Y$ and the number of different numeraires under which $Y$ is most similar to $X$. It can be easily noticed that similar results to the previous discussion regarding the USD, EUR and Middle Eastern cluster hold. Besides this, the performed analysis is able to uncover novel information about the correlation structure. Concretely, here we observe the similarity between Latin American BLR, CLP, COP and MXN, as well as between certain currencies from Asia. From the studied precious metals, the silver and platinum seem to exhibit similar dynamics to that of the gold, while the gold remains most similar to JPY. Moreover, we notice that GBP appears most similar to SGD instead of any other European currency, even when accounting for every possible numeraire from Europe, as compared to the SGD alone. Finally, we remark that HKD and SGD are the two assets which are most similar to the others. While the former currency may appear as a central regional trade unit, due to the fixed exchange rate to the USD, the later has a bigger role in global trading.

\begin{table}[h!]
  \begin{center}
    \caption{\textbf{Most similar assets.} The most similar peer (second and fifth column) to certain asset (first and fourth column) is the one which has strongest correlation to the latter in largest number of different numeraires. The number of such occurrences for each pair is given in the third and sixth column.}
   \label{tab:Close_assets}
    \begin{tabular}{|c|c|c||c|c|c|} % <-- Alignments: 1st column left, 2nd middle and 3rd right, with vertical lines in between
 \hline      
 Asset & Sim. asset & No. occur. & Asset & Sim. asset & No. occur. \\
     \hline
    AED & SAR & 51 & MAD & DKK & 28 \\ \hline
    ARS & USD & 43 & MXN & CLP & 29 \\ \hline
    AUD & NZD & 47 & MYR & SGD & 37 \\ \hline
    BHD & AED & 51 & NOK & SEK & 46 \\ \hline
    BRL & MXN & 50 & NZD & AUD & 51 \\ \hline
    CAD & SGD & 36 & PEN & HKD & 43 \\ \hline
    CHF & EUR & 46 & PHP & HKD & 38 \\ \hline
    CLP & COP & 17 & PLN & HUF & 51 \\ \hline
    CNY & HKD & 50 & RUB & COP & 38 \\ \hline
    COP & CLP & 50 & SAR & AED & 51 \\ \hline
    CZK & DKK & 51 & SEK & EUR & 47 \\ \hline
    DKK & EUR & 51 & SGD & TWD & 47 \\ \hline
    EUR & DKK & 51 & THB & HKD & 40 \\ \hline
    GBP & SGD & 27 & TND & MAD & 28 \\ \hline
    HKD & USD & 51 & TRY & SGD,ZAR & 19 \\ \hline
    HNL & USD & 51 & TWD & SGD & 44 \\ \hline
    HRK & DKK & 51 & USD & HKD & 51 \\ \hline
    HUF & PLN & 51 & UYU & AED & 51 \\ \hline
    IDR & MYR & 31 & VND & HKD & 50 \\ \hline
    ILS & SGD & 23 & XAG & XAU & 45 \\ \hline
    INR & HKD & 40 & XAU & JPY & 43 \\ \hline
    ISK & DKK & 49 & XCB & XCT & 51 \\ \hline
    JMD & USD & 51 & XCD & USD & 50 \\ \hline
    JPY & SGD & 30 & XCT & XCB & 51 \\ \hline
    KRW & TWD & 51 & XPT & XAU & 50 \\ \hline
    KWD & AED & 50 & ZAR & MXN & 48 \\ \hline
    LKR & USD & 51 & \multicolumn{3}{|c|}{} \\ \hline
    \end{tabular}
    \raggedright{%Note: For each asset its most similar asset is considered the one that has highest number of occurrences as the the most correlated with it in different numeraires. The number of such occurrences is also given in order to approximate the similarity.
    }
  \end{center}
\end{table}

\subsection{Network of partial correlations}

%The partial correlations between pairs of assets were obtained from the precision matrix by using (\ref{eq:Partial_corr}), which is calculated as inverse of the correlation matrix. We have kept only the statistically significant partial correlations with $p$ value less than $0.05/(53\cdot 53)$. 

Using the same data, we estimated the partial correlation between the assets, and as described in the Data section, we used the statistically significant relationships to build a network which describes their relatedness. 

\begin{figure*}[t!]
\begin{adjustwidth}{-0in}{0in}
\includegraphics[width=15cm]{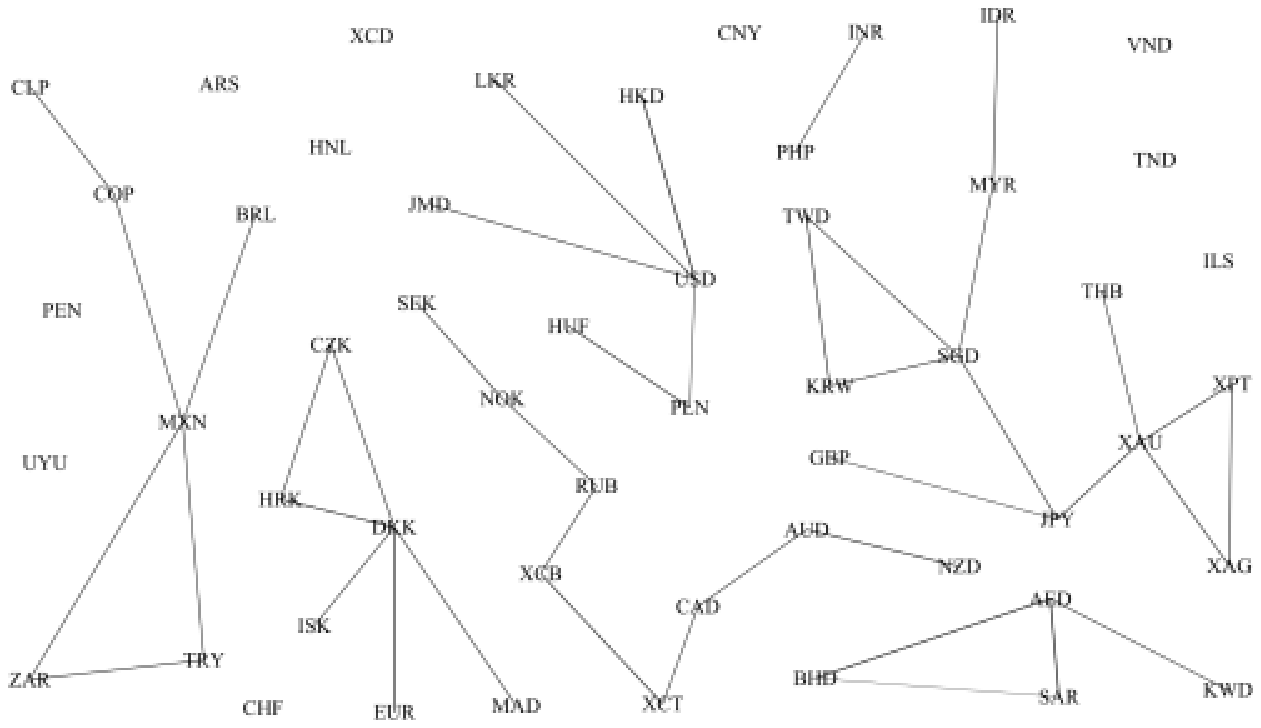}
\caption{\textbf{Partial correlation network.} The edges represent statistically significant partial correlations.
\label{fig:part_corr}}
\end{adjustwidth}
\end{figure*}

The statistically significant partial correlations are given in Table~\ref{tab:Part_corr}, while the resulting network is illustrated in Fig.~\ref{fig:part_corr}. The network is largely disconnected with only several isolated clusters being formed which can be explained by considering the geographical proximity of countries again. We note that similar findings about the relations between currencies of neighboring countries have been previously obtained with the MST technique~\cite{mizuno2006correlation, keskin2011topology}.

In particular, the largest cluster consists mainly of currencies from South and East Asia. In addition, precious metals can be found in this cluster due to their relation with JPY and THB. Here, it is also noteworthy to point out that the GBP exhibits negative partial correlation with JPY. Another easily interpretable cluster is the group of interconnected European currencies consisting of CZK, DKK, EUR, HRK, ISK and MAD. Interestingly, other European currencies, specifically PLN and HUF, are instead connected to a USD-based cluster due to the negative partial correlation between PLN and USD. This cluster further includes HKD, LKR and JML. The next cluster consists of the Middle east currencies AED, KWD and SAR, which are acompanied with BHD due to its rather strong negative partial correlation with AED. 

A rather intriguing cluster centred around MXN appears. This cluster, on the one hand, consists of neighboring Latin American currencies BRL, COP and CLP, and the geographically distant currencies ZAR nad TRY, on the other hand. The last cluster describes the relationships between oiled-based assets, mainly due to the price of West Texas Intermediate being partially correlated with the CAD, whereas the Brent Crude oil is related with RUB. Moreover, the oil production industry is probably the cause for the positive partial correlation between RUB and NOK, which in turn puts the latter with the neighboring SEK into this group, rather than into the EUR-based one. We point out that NOK and SEK have more strong unconditional correlations with the currencies from the other other EU countries than with the RUB. In this cluster the Oceania pair of AUD and NZD is included due to the significant partial correlation between AUD and CAD. Finally, INR and PHP form an isolated pair of mutually influential currencies from neighboring countries.

\begin{table}[h!]
  \begin{center}
    \caption{\textbf{Statistically significant partial correlations.}Note that the negative correlations are in bold.}
   \label{tab:Part_corr}
    \begin{tabular}{|l|l||l|l||l|l|}
 \hline      
 Pair & $\rho_{x,y}$ & Pair & $\rho_{x,y}$ & Pair & $\rho_{x,y}$ \\
     \hline
AED/SAR & 0.964 & CZK/DKK & 0.246 & AUD/CAD & 0.167 \\ \hline
HKD/USD & 0.936 & JPY/SGD & 0.245 & LKR/USD & 0.166 \\ \hline
AED/BHD & 0.795 & IDR/MYR & 0.242 & JMD/USD & 0.165 \\ \hline
DKK/EUR & 0.640 & SGD/TWD & 0.225 & DKK/ISK & 0.164 \\ \hline
XCB/XCT & 0.495 & DKK/MAD & 0.223 & MXN/TRY & 0.161 \\ \hline
HUF/PLN & 0.446 & CLP/COP & 0.221 & COP/MXN & 0.160 \\ \hline
AUD/NZD & 0.434 & MYR/SGD & 0.211 & AED/KWD & 0.145 \\ \hline
DKK/HRK & 0.434 & CZK/HRK & 0.208 & INR/PHP & 0.143 \\ \hline
KRW/TWD & 0.403 & MXN/ZAR & 0.203 & BRL/MXN & 0.139 \\ \hline
JPY/XAU & 0.362 & TRY/ZAR & 0.203 & THB/XAU & 0.139 \\ \hline
XAU/XPT & 0.347 & KRW/SGD & 0.192 & GBP/JPY & \textbf{-0.144} \\ \hline
XAG/XAU & 0.338 & NOK/RUB & 0.179 & PLN/USD & \textbf{-0.147} \\ \hline
NOK/SEK & 0.313 & RUB/XCB & 0.177 & BHD/SAR & \textbf{-0.807} \\ \hline
XAG/XPT & 0.280 & CAD/XCT & 0.174 & \multicolumn{2}{|c|}{} \\ \hline
    \end{tabular}
  \end{center}
\end{table}

\section{\label{sec:Conclusions}Conclusions}

In short summary, in this paper we investigated how the means, covariances and correlations of the logarithmic returns of assets are modified when the numeraire in a foreign exchange is changed. We also showed that the same techniques can be applied for the portfolio of an investor, and thus identify the most appropriate numeraire for him/her to use. This turned out to be the domestic currency of the investor. Finally, we showed that while the magnitude of the ordinary correlations are highly dependent on the choice of numeraire, the partial correlations are invariant in this aspect. The empirial analysis easily confirmed these findings.

%that the same transformations can be applied for portfolio as well, and that from investor's point of view the most appropriate numeraire for estimation of the associated risk is his or her domestic currency. We have finally shown that although correlations can be rather different when one substitutes one base currency with another, the partial correlations are invariant. The empirical studies have shown that the numeraire can have rather distorting effect on the correlation matrix by shadowing the correlations of the assets that have rather correlated dynamics with it. It was obtained that the partial correlations as numeraire invariant are immune to this effect.

\section*{Acknowledgement}

We are grateful to Pacific Exchange Rate Service for freely sharing their data. This research was partially supported by the Faculty of Computer Science and Engineering at ``SS. Cyril and Methodius'' University in Skopje, Macedonia and in part by DFG through grant ``Random search processes, L\'evy flights, and random walks on complex networks''.

\section*{Appendix} \label{sec:Appendix}

\begin{longtable}{||l|l||l|l||l|l||l|l||}
  %\begin{center}
    \caption{Correlation coefficients larger than
    0.6 when the numeraire is USD}
      
   \label{tab:Corr_USD}
    \\
       \hline  \\
        Pair & $r_{x,y}$ & Pair & $r_{x,y}$ & Pair & $r_{x,y}$ & Pair & $r_{x,y}$\\
      \hline  

AED/SAR & 0.985 & CZK/PLN & 0.836 & EUR/ISK & 0.761 & NZD/SGD & 0.644 \\ \hline
DKK/EUR & 0.980 & CZK/HUF & 0.834 & HRK/SEK & 0.758 & NOK/PLN & 0.639 \\ \hline
DKK/HRK & 0.974 & DKK/PLN & 0.834 & CZK/SEK & 0.755 & EUR/NOK & 0.635 \\ \hline
CZK/DKK & 0.958 & HRK/HUF & 0.828 & ISK/MAD & 0.746 & SGD/THB & 0.633 \\ \hline
EUR/HRK & 0.957 & HRK/PLN & 0.825 & KRW/SGD & 0.737 & AUD/CAD & 0.632 \\ \hline
CZK/HRK & 0.948 & EUR/HUF & 0.823 & AUD/NZD & 0.732 & DKK/NOK & 0.627 \\ \hline
CZK/EUR & 0.938 & EUR/PLN & 0.818 & MAD/SEK & 0.722 & HUF/NOK & 0.627 \\ \hline
DKK/MAD & 0.933 & DKK/ISK & 0.790 & SGD/TWD & 0.721 & CZK/NOK & 0.624 \\ \hline
EUR/MAD & 0.919 & MAD/PLN & 0.789 & HUF/SEK & 0.695 & ISK/SEK & 0.621 \\ \hline
AED/KWD & 0.918 & HUF/MAD & 0.789 & ISK/PLN & 0.693 & HRK/NOK & 0.620 \\ \hline
HRK/MAD & 0.915 & EUR/SEK & 0.783 & NOK/SEK & 0.692 & HRK/TND & 0.613 \\ \hline
KWD/SAR & 0.908 & HRK/ISK & 0.781 & PLN/SEK & 0.689 & DKK/TND & 0.611 \\ \hline
CZK/MAD & 0.905 & CZK/ISK & 0.780 & HUF/ISK & 0.686 & EUR/TND & 0.608 \\ \hline
HUF/PLN & 0.866 & DKK/SEK & 0.776 & AUD/SGD & 0.668 & & \\ \hline
DKK/HUF & 0.837 & KRW/TWD & 0.767 & XCB/XCT & 0.647 & & \\ \hline

    %\end{tabular}
  %\end{center}
\end{longtable}

\begin{longtable}{||l|l||l|l||l|l||l|l||}
    \caption{Correlation coefficients larger than 0.6 when the numeraire is EUR}
   \label{tab:Corr_EUR}
\\
       \hline \\
        Pair & $r_{x,y}$ & Pair & $r_{x,y}$ & Pair & $r_{x,y}$ & Pair & $r_{x,y}$\\
      \hline  
HKD/USD & 0.998 & PEN/THB & 0.804 & KWD/THB & 0.726 & ILS/JMD & 0.647 \\ \hline
AED/SAR & 0.995 & AED/JMD & 0.802 & CNY/HNL & 0.725 & ILS/SGD & 0.646 \\ \hline
AED/KWD & 0.970 & JMD/SAR & 0.802 & AED/PHP & 0.724 & IDR/JMD & 0.646 \\ \hline
KWD/SAR & 0.967 & PEN/PHP & 0.800 & PHP/SAR & 0.722 & CAD/THB & 0.645 \\ \hline
LKR/USD & 0.954 & KRW/TWD & 0.800 & AUD/NZD & 0.720 & ILS/THB & 0.645 \\ \hline
HKD/LKR & 0.952 & THB/XCD & 0.798 & INR/MYR & 0.718 & AED/HNL & 0.644 \\ \hline
JMD/USD & 0.949 & INR/SGD & 0.796 & HKD/MYR & 0.717 & MAD/TWD & 0.644 \\ \hline
CNY/HKD & 0.948 & MAD/USD & 0.795 & MYR/USD & 0.711 & IDR/LKR & 0.644 \\ \hline
HKD/JMD & 0.947 & CNY/SGD & 0.795 & MAD/THB & 0.707 & HNL/SAR & 0.642 \\ \hline
CNY/USD & 0.946 & HKD/MAD & 0.794 & KWD/PEN & 0.706 & COP/MXN & 0.642 \\ \hline
JMD/LKR & 0.907 & INR/PEN & 0.793 & CAD/SGD & 0.706 & ILS/PHP & 0.640 \\ \hline
USD/XCD & 0.906 & KWD/USD & 0.793 & INR/KWD & 0.706 & CLP/PHP & 0.640 \\ \hline
HKD/XCD & 0.905 & HKD/KWD & 0.793 & SAR/TWD & 0.703 & AED/MYR & 0.638 \\ \hline
CNY/LKR & 0.905 & AED/LKR & 0.792 & SGD/XCD & 0.703 & ILS/LKR & 0.638 \\ \hline
CNY/JMD & 0.902 & LKR/SAR & 0.790 & AED/TWD & 0.703 & KRW/MYR & 0.636 \\ \hline
HKD/THB & 0.881 & PEN/TWD & 0.781 & SAR/SGD & 0.693 & MYR/SAR & 0.636 \\ \hline
SGD/TWD & 0.881 & KRW/SGD & 0.781 & AED/SGD & 0.693 & CAD/PEN & 0.636 \\ \hline
THB/USD & 0.878 & HKD/SGD & 0.780 & MAD/PHP & 0.691 & HKD/VND & 0.635 \\ \hline
JMD/XCD & 0.875 & JMD/TWD & 0.778 & HNL/XCD & 0.690 & CAD/PHP & 0.635 \\ \hline
LKR/XCD & 0.869 & PEN/XCD & 0.776 & LKR/MYR & 0.689 & USD/VND & 0.635 \\ \hline
CNY/THB & 0.868 & SGD/USD & 0.773 & KWD/PHP & 0.689 & IDR/SGD & 0.634 \\ \hline
CNY/XCD & 0.863 & LKR/TWD & 0.773 & JMD/MYR & 0.686 & ILS/TWD & 0.632 \\ \hline
HKD/PHP & 0.862 & MYR/SGD & 0.771 & MYR/PEN & 0.685 & CAD/INR & 0.632 \\ \hline
HKD/PEN & 0.859 & AED/XCD & 0.770 & KRW/THB & 0.683 & KWD/MAD & 0.631 \\ \hline
PHP/USD & 0.858 & SAR/XCD & 0.770 & IDR/MYR & 0.682 & IDR/TWD & 0.631 \\ \hline
PEN/USD & 0.857 & JMD/KWD & 0.767 & MAD/PEN & 0.679 & HNL/PHP & 0.629 \\ \hline
THB/TWD & 0.855 & PHP/XCD & 0.767 & KRW/PHP & 0.677 & AED/CAD & 0.628 \\ \hline
PHP/THB & 0.852 & CNY/KWD & 0.765 & CLP/TWD & 0.676 & ILS/PEN & 0.627 \\ \hline
HKD/INR & 0.848 & INR/XCD & 0.764 & ILS/USD & 0.674 & CAD/SAR & 0.627 \\ \hline
SGD/THB & 0.847 & HNL/USD & 0.763 & HKD/ILS & 0.673 & ILS/INR & 0.625 \\ \hline
PHP/TWD & 0.846 & HKD/HNL & 0.762 & CNY/IDR & 0.673 & CAD/CLP & 0.624 \\ \hline
INR/USD & 0.845 & AED/THB & 0.761 & CLP/PEN & 0.672 & DKK/HRK & 0.624 \\ \hline
LKR/THB & 0.843 & SAR/THB & 0.761 & KWD/SGD & 0.671 & CLP/THB & 0.622 \\ \hline
JMD/THB & 0.842 & LKR/MAD & 0.760 & KWD/TWD & 0.669 & CAD/KWD & 0.621 \\ \hline
INR/PHP & 0.837 & PEN/SGD & 0.759 & IDR/PHP & 0.669 & ILS/XCD & 0.620 \\ \hline
CNY/PHP & 0.837 & MYR/THB & 0.758 & INR/KRW & 0.669 & CAD/MXN & 0.619 \\ \hline
AED/HKD & 0.837 & JMD/MAD & 0.757 & HNL/THB & 0.669 & IDR/PEN & 0.618 \\ \hline
INR/THB & 0.837 & MYR/TWD & 0.757 & CAD/TWD & 0.668 & HKD/UYU & 0.617 \\ \hline
AED/USD & 0.836 & JMD/SGD & 0.754 & CLP/SGD & 0.668 & USD/UYU & 0.616 \\ \hline
HKD/SAR & 0.836 & KWD/LKR & 0.750 & HKD/IDR & 0.667 & MAD/SGD & 0.613 \\ \hline
SAR/USD & 0.836 & AED/UYU & 0.747 & IDR/THB & 0.666 & HNL/TWD & 0.613 \\ \hline
CNY/INR & 0.835 & CNY/MAD & 0.745 & IDR/USD & 0.665 & KWD/MYR & 0.611 \\ \hline
CNY/PEN & 0.834 & SAR/UYU & 0.743 & CLP/COP & 0.665 & CAD/HKD & 0.608 \\ \hline
INR/TWD & 0.829 & INR/SAR & 0.742 & AUD/CAD & 0.665 & IDR/XCD & 0.608 \\ \hline
PHP/SGD & 0.826 & PEN/SAR & 0.741 & AUD/SGD & 0.661 & CAD/CNY & 0.607 \\ \hline
CNY/TWD & 0.822 & AED/PEN & 0.739 & XCB/XCT & 0.660 & HNL/KWD & 0.605 \\ \hline
JMD/PEN & 0.820 & AED/INR & 0.739 & INR/MAD & 0.659 & MXN/PEN & 0.604 \\ \hline
JMD/PHP & 0.820 & MYR/PHP & 0.738 & IDR/INR & 0.658 & AUD/TWD & 0.604 \\ \hline
LKR/PHP & 0.819 & CNY/MYR & 0.736 & CNY/ILS & 0.657 & CLP/KRW & 0.604 \\ \hline
HKD/TWD & 0.816 & LKR/SGD & 0.736 & MYR/XCD & 0.657 & JMD/VND & 0.604 \\ \hline
INR/LKR & 0.815 & KWD/XCD & 0.731 & HNL/PEN & 0.657 & LKR/VND & 0.604 \\ \hline
INR/JMD & 0.815 & TWD/XCD & 0.729 & HNL/INR & 0.652 & MXN/SGD & 0.603 \\ \hline
LKR/PEN & 0.814 & KWD/UYU & 0.727 & MAD/SAR & 0.650 & AUD/KRW & 0.603 \\ \hline
TWD/USD & 0.810 & HNL/JMD & 0.727 & CLP/INR & 0.650 & CNY/VND & 0.603 \\ \hline
AED/CNY & 0.808 & MAD/XCD & 0.726 & AED/MAD & 0.649 & CAD/USD & 0.602 \\ \hline
CNY/SAR & 0.808 & HNL/LKR & 0.726 & CLP/MXN & 0.647 &  &  \\ \hline
%\end{center}
%\end{tabular}
\end{longtable}

\begin{longtable}{||l|l||l|l||l|l||l|l||}
  %\begin{center}
    \caption{Correlation coefficients larger than 0.6 when the numeraire is SAR}
   \label{tab:Corr_SAR}
\\
       \hline \\
        Pair & $r_{x,y}$ & Pair & $r_{x,y}$ & Pair & $r_{x,y}$ & Pair & $r_{x,y}$\\
      \hline 
     HKD/USD & 0.995 & EUR/SEK & 0.799 & JMD/XCD & 0.702 & CNY/INR & 0.645 \\ \hline
DKK/EUR & 0.983 & DKK/SEK & 0.795 & NOK/SEK & 0.701 & JMD/THB & 0.645 \\ \hline
DKK/HRK & 0.977 & EUR/ISK & 0.793 & HKD/PEN & 0.699 & HUF/NOK & 0.644 \\ \hline
CZK/DKK & 0.962 & MAD/PLN & 0.792 & CNY/PHP & 0.696 & CZK/NOK & 0.642 \\ \hline
EUR/HRK & 0.962 & HUF/MAD & 0.791 & PEN/USD & 0.695 & PEN/PHP & 0.641 \\ \hline
CZK/HRK & 0.954 & ISK/MAD & 0.785 & LKR/XCD & 0.694 & DKK/SGD & 0.641 \\ \hline
CZK/EUR & 0.945 & HRK/SEK & 0.779 & INR/PHP & 0.692 & TWD/USD & 0.640 \\ \hline
DKK/MAD & 0.926 & KRW/TWD & 0.777 & INR/TWD & 0.686 & HRK/NOK & 0.638 \\ \hline
EUR/MAD & 0.912 & CZK/SEK & 0.776 & MAD/SGD & 0.683 & ILS/SGD & 0.637 \\ \hline
HRK/MAD & 0.910 & JMD/LKR & 0.773 & CNY/TWD & 0.682 & AUD/CAD & 0.633 \\ \hline
CZK/MAD & 0.899 & CNY/LKR & 0.772 & CNY/XCD & 0.674 & PLN/SGD & 0.633 \\ \hline
LKR/USD & 0.888 & SGD/THB & 0.772 & INR/THB & 0.671 & ILS/MAD & 0.632 \\ \hline
HKD/LKR & 0.884 & USD/XCD & 0.771 & LKR/THB & 0.666 & JMD/PHP & 0.632 \\ \hline
HUF/PLN & 0.876 & HKD/XCD & 0.768 & HKD/INR & 0.664 & HRK/SGD & 0.630 \\ \hline
CNY/HKD & 0.866 & THB/TWD & 0.764 & DKK/TND & 0.663 & KRW/THB & 0.628 \\ \hline
JMD/USD & 0.864 & KRW/SGD & 0.760 & HRK/TND & 0.663 & PEN/THB & 0.628 \\ \hline
CNY/USD & 0.862 & PHP/TWD & 0.756 & ISK/SEK & 0.661 & CZK/SGD & 0.623 \\ \hline
HKD/JMD & 0.860 & CNY/JMD & 0.750 & AUD/SGD & 0.661 & INR/SGD & 0.623 \\ \hline
DKK/HUF & 0.851 & PHP/THB & 0.748 & EUR/TND & 0.658 & NZD/SGD & 0.621 \\ \hline
CZK/PLN & 0.850 & PHP/SGD & 0.743 & XCB/XCT & 0.657 & HUF/SGD & 0.619 \\ \hline
DKK/PLN & 0.848 & HKD/THB & 0.741 & CNY/PEN & 0.656 & PEN/TWD & 0.619 \\ \hline
CZK/HUF & 0.848 & MAD/SEK & 0.739 & NOK/PLN & 0.656 & KRW/PHP & 0.617 \\ \hline
HRK/HUF & 0.842 & HKD/PHP & 0.735 & CNY/SGD & 0.656 & LKR/PEN & 0.612 \\ \hline
HRK/PLN & 0.840 & THB/USD & 0.732 & HKD/TWD & 0.655 & JMD/PEN & 0.612 \\ \hline
EUR/HUF & 0.838 & AUD/NZD & 0.731 & INR/USD & 0.654 & CLP/COP & 0.609 \\ \hline
EUR/PLN & 0.834 & CNY/THB & 0.730 & MAD/TND & 0.652 & MYR/SGD & 0.605 \\ \hline
AED/BHD & 0.821 & ISK/PLN & 0.725 & CZK/TND & 0.651 & CAD/SGD & 0.604 \\ \hline
DKK/ISK & 0.820 & PHP/USD & 0.723 & EUR/NOK & 0.650 & INR/LKR & 0.604 \\ \hline
SGD/TWD & 0.820 & HUF/ISK & 0.718 & EUR/SGD & 0.647 & HKD/SGD & 0.603 \\ \hline
HRK/ISK & 0.810 & HUF/SEK & 0.717 & DKK/NOK & 0.645 & MYR/TWD & 0.603 \\ \hline
CZK/ISK & 0.809 & PLN/SEK & 0.712 & LKR/PHP & 0.645 & MAD/THB & 0.601 \\ \hline
    %\end{tabular}
  %\end{center}
\end{longtable}

%\bibliographystyle{unsrt}
%\bibliography{forex}

\end{document}